\documentclass[letterpaper, 10 pt, conference]{ieeeconf}

\IEEEoverridecommandlockouts                              

\makeatletter
\newcommand{\ftype@noticebox}{8}
\newcommand{\notice}{\enlargethispage{2\baselineskip}\@float{noticebox}[b]\footnotesize\@noticestring \end@float }

\newcommand{\@noticestring}{2023 IEEE International Conference on Robotics and Automation (ICRA)
}
\makeatother

\usepackage{balance}

\newcommand{\add}[1]{#1}

\usepackage[utf8]{inputenc} \usepackage[T1]{fontenc}    \usepackage{url}            

\usepackage{booktabs}       \usepackage{caption}

\usepackage{amsfonts}       \usepackage{nicefrac}       \usepackage{microtype}      \usepackage{xcolor}         \usepackage{graphicx}

\usepackage{amsthm}

\usepackage{amsmath}
\usepackage{mathtools}
\usepackage{tabularx}
\usepackage{framed}
\usepackage{bbm}
\usepackage{bm}
\usepackage{xspace}
\usepackage{amssymb}
\usepackage{multirow}
\usepackage{physics}
\usepackage{siunitx}
\usepackage{algorithm}
\usepackage{algorithmic}

\usepackage{cite}

\usepackage{nccmath}

\usepackage{colortbl}

\usepackage{placeins}

\usepackage{subcaption}

\usepackage{lipsum}

\usepackage{thmtools}
\usepackage{thm-restate}

\makeatletter
\let\NAT@parse\undefined
\makeatother
\usepackage{hyperref}       \hypersetup{
    colorlinks,
    linkcolor={blue!90!black},
    citecolor={blue!90!black},
urlcolor={blue!95!black}
}

\usepackage[capitalise]{cleveref}       \usepackage{crossreftools}  \usepackage[all=normal,paragraphs=tight,floats=tight,mathspacing=tight,wordspacing=tight,tracking=tight]{savetrees} \newcommand*{\Rb}{\mathbb{R}}

\DeclareMathOperator*{\Eb}{\mathbb{E}}

\DeclareMathOperator*{\SM}{\sigma_{\textrm{SM}}}

\def\vu{{\vb{u}}}
\def\vx{{\vb{x}}}

\DeclareBoldMathCommand\vpi{\pi}

\def\vU{{\vb{U}}}
\def\vX{{\vb{X}}} \newcommand*\numx{n_{\vx}}
\newcommand*\numu{n_{\vu}}

\newcommand*\negi{{\neg i}}

\newcommand*\ui{u^{i}}

\newcommand*\pii{\pi^{i}}

\newcommand*\pinegi{\vpi^{\negi}}

\newcommand{\one}{{\mathfrak{1}}}
\newcommand{\two}{{\mathfrak{2}}}

\newcommand{\notone}{{\neg \mathfrak{1}}}

\newcommand*\ttNoInf{{\texttt{NoInf}}}
\newcommand*\ttYield{{\texttt{Yield}}}
\newcommand*\ttNoYield{{\texttt{NoYield}}}
\newcommand*\ttML{{\texttt{ML}}}
\newcommand*\ttQMDP{{\texttt{MPOGames}}} \definecolor{palegreen}{HTML}{F1F8E9}
\definecolor{palered}{HTML}{FFEBEE}
\definecolor{green600}{HTML}{43A047}
\definecolor{figgold}{HTML}{a29120}
\definecolor{figred}{HTML}{f5494e}

\definecolor{red1}{HTML}{fc4f30}
\definecolor{blue1}{HTML}{068dd4} 

\newtheorem*{remark}{Remark}

\makeatletter
\def\footnoterule{\kern-3\p@
  \hrule \@width 2in \kern 2.6\p@} \makeatother

\DeclareRobustCommand*{\numauthorrefmark}[1]{\raisebox{0pt}[0pt][0pt]{\textsuperscript{\footnotesize\ensuremath{#1}}}}

\title{\LARGE \bf MPOGames: Efficient Multimodal Partially Observable Dynamic Games}
\author{\authorblockN{Oswin So\numauthorrefmark{1,2,*}, Paul Drews\numauthorrefmark{2}, Thomas Balch\numauthorrefmark{2}, Velin Dimitrov\numauthorrefmark{2}, Guy Rosman\numauthorrefmark{2}, Evangelos A. Theodorou\numauthorrefmark{1}\thanks{\numauthorrefmark{1} Georgia Institute of Technology.}
\thanks{\numauthorrefmark{2} Toyota Research Institute.}
\thanks{\numauthorrefmark{*} Corresponding Author. \href{mailto:oswinso@gatech.edu}{oswinso@gatech.edu}.}
\thanks{\newline This work was supported by the Toyota Research Institute (TRI). This article solely reflects the opinions and conclusions of its authors and not TRI or any other Toyota entity. Their support is gratefully acknowledged.}
}}

\begin{document}

\maketitle
\notice
\thispagestyle{empty}
\pagestyle{empty}

\global\csname @topnum\endcsname 0
\global\csname @botnum\endcsname 0
\begin{abstract}
Game theoretic methods have become popular for planning and prediction in situations involving rich multi-agent interactions.
However, these methods often assume the existence of a single local Nash equilibria and are hence unable to handle uncertainty in the intentions of different agents.
While maximum entropy (MaxEnt) dynamic games try to address this issue, practical approaches solve for MaxEnt Nash equilibria using linear-quadratic approximations which are restricted to \textit{unimodal} responses and unsuitable for scenarios with multiple local Nash equilibria.
By reformulating the problem as a POMDP, we propose MPOGames, a method for efficiently solving MaxEnt dynamic games that captures the interactions between local Nash equilibria. We show the importance of uncertainty-aware game theoretic methods via a two-agent merge case study.
Finally, we prove the real-time capabilities of our approach with hardware experiments on a 1/10th scale car platform.
\end{abstract}

\begin{figure}
    \centering
    \includegraphics[width=0.999\linewidth]{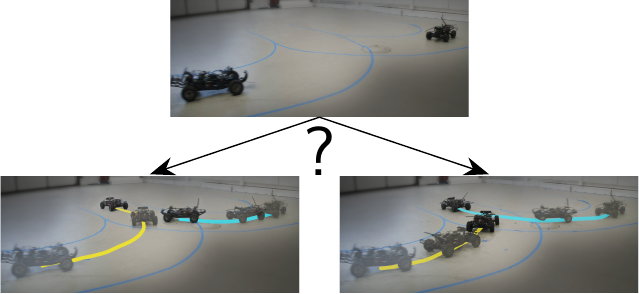}
    \caption{
        The ego agent (left) seeks to merge into a lane safely without collisions.
With multiple local minima present, MPOGames \textit{predicts} likely plans for other agents, \textit{infers} their probabilities, then \textit{hedges} against all possibilities.
        Methods that fail to do so may result in catastrophic collisions.}
    \label{fig:hero}
    \vspace{-\baselineskip}
\end{figure} 

\section{Introduction and Related Work}

In recent years, robots have been seen increasing use in applications with multiple interacting agents.
In these applications, actions taken by the robot cannot be considered in isolation and must take the responses of other agents into account, especially in areas such as autonomous driving \cite{sadigh2018planning}, robotic arms \cite{amor2014interaction} and surgical robotics \cite{zhou2018early} where failures can be catastrophic.
For example, in the autonomous driving setting, two agents at a merge must negotiate and agree on which agent merges first (\cref{fig:hero}). Failure to consider the responses of the other agent could result in serious collisions.

Game-theoretic (GT) approaches have become increasingly popular for planning in multi-agent scenarios \cite{cleac2019algames,fridovich2020efficient,laine2021multi,mehr2021maximum,schwarting2021stochastic,so2022multimodal,wang2019game,wang2020multi,peters2020inference}. By assuming that agents act rationally, these approaches decouple the problem of planning and prediction into the problems of finding objective functions for each agent and solving for Nash equilibrium of the resulting non-cooperative game. This provides a more principled and interpretable alternative to the prediction of other agents when compared to black-box approaches that use deep neural networks \cite{salzmann2020trajectron++,ban2022deep}.

In practice, assuming rational agents is too strict. Humans face cognitive limitations and make irrational decisions that are satisfactory but suboptimal, an idea known as ``bounded rationality'' or ``noisy rationality'' \cite{luce1960individual,simon1972theories}. Even autonomous agents are rarely \textit{exactly} optimal and often make suboptimal decisions due to finite computational resources.
One method of addressing this disconnect is the Maximum Entropy (MaxEnt) framework, which has been applied to diverse areas such as inverse reinforcement learning \cite{ziebart2008maximum,mehr2021maximum}, forecasting \cite{evans2021maximum} and biology \cite{de2018introduction}. In particular, this combination of MaxEnt and GT has been proposed for learning objectives for agents from a traffic dataset \cite{mehr2021maximum}. However, since the exact solution is computationally intractable, the authors use a linear quadratic (LQ) approximation of the game for which a closed-form solution can be efficiently obtained.
The LQ approximation, popular among recent approaches (e.g., \cite{fridovich2020efficient,mehr2021maximum}),
only considers a single local minimum and consequently produces \textit{unimodal} predictions for each agent. This fails to capture complex uncertainty-dependent hedging behaviors which mediate between other agents' possible plans in the true solution of the MaxEnt game.

These hedging behaviors are
closely related to partially observable Markov decision process (POMDP) in the single-player setting, where solutions must consider the \textit{distribution} of states when solving for the optimal controls. Similar multi-agent scenarios have been considered via POMDPs in \cite{qiu2020latent,hu2022active, hu2022sharp}. Also related are partially observable stochastic games 
which are usually intractable \cite{sadigh2018planning} and mostly limited to the discrete setting \cite{emery-montemerloApproximateSolutions2004,hansenDynamicProgramming2004,horakSolvingPartially2019,kumarDynamicProgramming2009}.

One big challenge for methods that seek to address partial observability and multimodality is the problem of computational efficiency and scalability (e.g., in the case of POMDPs). Furthermore, only a few of the existing GT methods perform experiments on hardware \cite{wang2019game,wang2020multi,kavuncu2021potential}, where the robustness of the method to noise and delays in the system is crucial.
To tackle both of these issues, we present MPOGames, a framework for efficiently solving multimodal MaxEnt dynamic games by reformulating the problem as a
game of incomplete information
and efficiently finding approximate solutions to the equivalent POMDP problem. We showcase the real-time potential of our method via extensive hardware comparisons.

Our contributions in this work are as follows.
\begin{enumerate}
\item A computationally efficient framework for GT planning in multi-agent scenarios that produces more accurate solutions in the presence of multiple local minima compared to previous works.
    \item A case-study in simulation comparing how GT methods handle multimodality that show the importance of inference and handling uncertainty.
    \item Analysis of the performance of GT methods in multimodal situations 
    with a 1/10th scale car platform.
    These results demonstrate the robustness and tractability of MPOGames on hardware.
\end{enumerate} 

\section{Preliminaries}
\subsection{Discrete Dynamic Games and Nash Equilibria}
We consider a discrete dynamic game with $N$ players.
Let $\vu_t = [u_t^1, \dots, u_t^N] = [u_t^i, \vu_t^\negi] \in \Rb^{\numu}$ denote the joint control vector for all players, where we use $(\cdot)^i$ and $(\cdot)^\negi$ to denote the partition of $(\cdot)$ into parts belonging to agent $i$ and other agents respectively.
We denote \add{by} $\vx_t \in \Rb^{\numx}$ the full state of the system at timestep $t$ with nonlinear dynamics
\begin{equation}
    \vx_{t+1} = f(\vx_t, u_t^1, \dots, u_t^N) = f(\vx_t, \vu_t).
\end{equation}
Time indices are dropped when they are clear from the context.
Each agent's objective is to minimize a corresponding finite-horizon cost function $J^i$ with running cost $l^i$ and terminal cost $\Phi^i$ with respect to the control trajectory $\vU = [\vu_1, \dots, \vu_{T-1}]$
\begin{equation} \label{eq:agent_obj_det}
    \min_{U^i \in \mathcal{U}^i} J^i(U^i, \vU^{\negi})
    \; =\;  \min_{U^i \in \mathcal{U}^i} \Phi^i(\vx_T) + \sum_{t=1}^{T-1} l^i(\vx_t, \vu_t),
\end{equation}
where $\mathcal{U}^i$ denotes the set of feasible control trajectories for agent $i$.
Note that the objective function $J^i$ is a function of both $U^i$ and $\vU^\negi$, but the optimization is only over $U^i$.
Hence, the quantity of interest here is the Nash Equilibrium (NE), i.e., find controls $\vU^*$ such that for each agent $i$,
\begin{equation} \label{eq:nash_def}
    J^i(U^{i*}, \vU^{\negi*}) \leq J^i(U^i, \vU^{\negi*}), \quad \forall U^i \in \mathcal{U}^i.
\end{equation}

\begin{figure*}
    \centering
\includegraphics[width=0.99\linewidth]{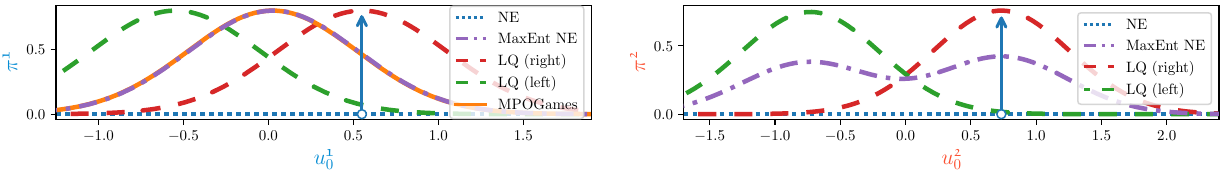}
\vspace{-0.2\baselineskip}
    \caption{Optimal policies for the ego agent P$\one$ (\textbf{left}) and non-ego agent P$\two$ (\textbf{right}) on the toy problem with $\epsilon=0.1$. For $\textcolor{blue1}{\pi^\one}$, only MaxEnt and POMDP hedges against both local NE of \textcolor{red1}{$\pi^\two$}. The deterministic policy of NE is denoted by the Dirac delta.
    }
    \label{fig:example}
    \vspace{-1.2\baselineskip}
\end{figure*} \subsection{Maximum Entropy Dynamic Games}
Agents rarely act \textit{exactly} optimally due to cognitive or computational limitations and settle for satisfactory decisions \cite{simon1972theories}.
We can model suboptimality via the noisy rationality model \add{\cite{luce1960individual}}, where controls are stochastic
according to the MaxEnt framework \cite{jaynes1982rationale,ziebart2008maximum}.
Consider a \textit{stochastic} control policy $\pii(\ui, \vx)$ and introduce an entropy regularization term to the original objective \eqref{eq:agent_obj_det}. We then consider the expected cost under all agents' policies $\vpi$
\begin{equation}
    J^i(\vpi) = \Eb_{\vu_t \sim \vpi} \left[ \Phi^i(\vx_T) + \sum_{t=1}^{T-1} l^i(\vx_t, \vu_t) - \frac{1}{\beta} H[\pii(\cdot | \vx_t)] \right],
\end{equation}
where $\beta > 0$ denotes inverse temperature or the rationality coefficient, and $H[\pi]$ is the Shannon entropy of $\pi$, defined as
\begin{equation}
    H[\pi] \coloneqq -\Eb_{u \sim \pi}[ \log \pi(u) ] = - \int \pi(u) \log \pi(u) \dd{u}.
\end{equation}
The resulting MaxEnt NE problem (referred to as ``Entropic Cost Equilibrium'' in \cite{mehr2021maximum}) is then to find stochastic policies $\vpi^*$ such that the following holds for each agent
\begin{equation} \label{eq:maxent_nash_def}
    J^i(\pi^{i*}, \pi^{\negi*}) \leq J^i(\pi^i, \pi^{\negi*}), \quad \forall \pi^i \in \Pi^i,
\end{equation}
for feasible control policies $\Pi^i$ from any initial state.
Let the value function $V^i$ for agent $i$ given the policies of other agents $\pinegi$ be
\begin{equation}
    V^i_{t}(\vx_t) \coloneqq \inf_{\pii}\{ \, J^i(\vx, \pi^i, \pinegi) \, \}.
\end{equation}
The optimal policy $\pi^{i*}$ takes the form \cite{mehr2021maximum,so2022multimodal}
\begin{equation} \label{eq:general_opt_pol}
    \pi^{i*}(\ui | \vx_t) = {Z^i(\vx_t)}^{-1} \exp\left( -\beta Q(\vx_t, \ui) \right),
\end{equation}
where
\begin{equation} \label{eq:maxent_q}
    Q^i(\vx_t, \ui_t) \coloneqq \Eb_{\pinegi}\left[{V^i}_{t+1}(\vx_{t+1}) + l^i(\vx_t, \vu)\right],
\end{equation}
and $Z^i$ denotes the partition function
\begin{equation} \label{eq:partition_fn}
    Z^i(\vx_t) \coloneqq \int \exp\left(
        -\beta \Eb_{\pinegi}\left[{V^i}_{t+1}(\vx_{t+1}) + l^i(\vx_t, \vu)\right]
    \right) \dd{\ui}.
\end{equation}
Then, the value function $V^i$ takes the form \cite[Appendix A]{so2021maximum}
\begin{equation} \label{eq:general_opt_valuefn}
    V^i_{t}(\vx_t) = - \ln Z^i(\vx_t).
\end{equation}
While this gives us an expression for $\pi^{i*}$ and $V^i$, it is generally intractable to compute the integral \eqref{eq:partition_fn} and solve for $\vpi^*$ and $Z^i$ under general nonlinear costs and dynamics.

One way of resolving this problem is to approximate the problem as a linear-quadratic (LQ) game \cite{mehr2021maximum, so2022multimodal} in a manner similar to iLQR \cite{li2004iterative}, an approach also taken in the iLQGames \cite{fridovich2020efficient,fridovich2020iterative} family of NE solvers. By doing so, the approximate game can be solved exactly \cite[Lemma 1 and Lemma 2]{so2022multimodal}.

\subsection{LQ is problematic for Multiodal MaxEnt NE}
While taking LQ approximations yields a computationally efficient closed-form simulation for both the original NE and the MaxEnt NE problem, this approximation can be especially problematic in the MaxEnt NE setting.
While the value function in the deterministic case does not depend on the cost function at states away from the NE, this is not the case for MaxEnt NE due to the integral in \eqref{eq:general_opt_pol} and \eqref{eq:general_opt_valuefn}.

\noindent\textbf{Illustrative Toy Example. }
We illustrate this using a toy example where the MaxEnt NE can be solved for in closed-form.
Consider the following two-agent single timestep dynamic game with
$\vx = [\textcolor{blue1}{x^\one}, \textcolor{red1}{x^\two}] \in \Rb^2$ and $\vu = [\textcolor{blue1}{u^\one}, \textcolor{red1}{u^\two}] \in \Rb^2$ with single-integrator dynamics and costs defined by
\begin{align}
    \textcolor{blue1}{J^\one} &= \frac{1}{2} ( \textcolor{blue1}{u_0^\one} )^2 + \frac{3}{2}( \textcolor{blue1}{x_1^\one} - \textcolor{red1}{x_1^\two} )^2, \\
    \textcolor{red1}{J^\two} &= \frac{1}{2} ( \textcolor{red1}{u_0^\two} )^2 + \Phi^\two( \textcolor{red1}{x_1^\two} ), \\
    \Phi^\two( \textcolor{red1}{x_1^\two} ) &=
        \SM\left(\frac{3}{2}( \textcolor{red1}{x_1^\two} - 1)^2, \; \frac{3}{2}( \textcolor{red1}{x_1^\two} + 1)^2+\epsilon \right), \label{eq:example:phi} \\
    \SM(a,b) &\coloneqq -\ln(e^{-a}+e^{-b}),
\end{align}
where $\epsilon > 0$ \add{controls the ``height'' of the left local minima (\cref{fig:example} top right)} and $\textcolor{blue1}{x_0^\one} = \textcolor{red1}{x_0^\two} = 0$. The $\SM$ in \eqref{eq:example:phi} acts as a smooth minimum of the two quadratic state costs.
In other words, \textcolor{red1}{P$\two$} wants to reach either $-1$ or $1$ (with a small preference for $1$), while \textcolor{blue1}{P$\one$} wants to track \textcolor{red1}{P$\two$}.

\noindent \textbf{NE: }
In the regular NE case, the optimization problem to solve for $\textcolor{blue1}{u_0^\one}$ is exactly quadratic and gives $\textcolor{blue1}{u_0^{\one*}} = \frac{3}{4} \textcolor{red1}{u_0^\two}$. Note that solution is \textit{independent} of $\epsilon$. We can also solve for $\textcolor{red1}{u_0^{\two*}}$ numerically to obtain $\textcolor{red1}{u_0^{\two*}} = \textcolor{red1}{x_1^{\two*}} \approx 0.73$. 

\noindent \textbf{Exact MaxEnt NE: }
When solving the MaxEnt NE case, $\textcolor{red1}{\pi^{\two*}}$ is a \textit{Gaussian mixture}.
Note the ``shape'' of the cost function $\textcolor{red1}{J^\two}$ determines the form of $\textcolor{red1}{\pi^\two}$.
Consequently, the optimal $\textcolor{blue1}{\pi^{\one*}}$ is Gaussian with mean $0.25 \tanh(\epsilon/2)$.
In comparison to the NE case, both $\textcolor{blue1}{\pi^{\one*}}$ and $\textcolor{red1}{\pi^{\two*}}$ depend on $\epsilon$.
Moreover, for $\epsilon \to 0$, the mean of $\textcolor{blue1}{\pi^{\one*}}$ approaches zero, which is \textit{qualitatively different} behavior to the previous case.

\noindent \textbf{LQ Approx MaxEnt NE: }
Using the LQ approximation, we obtain the following two sets of local LQ NE:
\begin{align*}
\textcolor{blue1}{\pi^{\one*, (1)}} &\approx \mathcal{N}(+0.55, 0.5), 
    & \textcolor{red1}{\pi^{\two*, (1)}} &\approx \mathcal{N}\left( +0.73, 0.53 \right), \\
\textcolor{blue1}{\pi^{\one*, (2)}} &\approx \mathcal{N}(-0.55, 0.5), 
    & \textcolor{red1}{\pi^{\two*, (2)}} &\approx \mathcal{N}\left( -0.73, 0.53 \right).
\end{align*}

For each method, we plot $\pi$ for both agents in \cref{fig:example}.
While the true MaxEnt $\textcolor{blue1}{\pi^\one}$ hedges against both modes of $\textcolor{red1}{\pi^\two}$ with near zero mean, neither LQ approximations capture this.

\section{Solving MaxEnt Dynamic Games as POMDPs}
\begin{figure*}
    \centering
    \includegraphics[width=\textwidth]{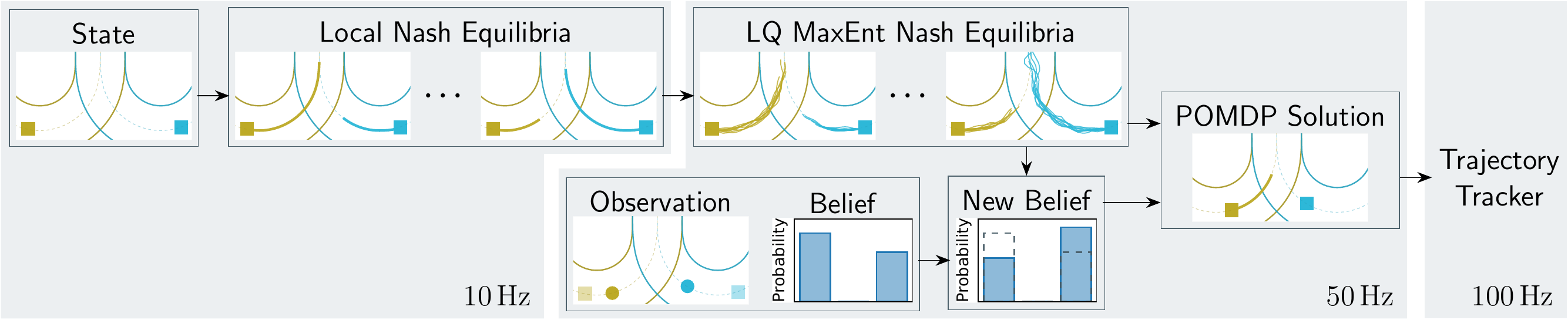}
    \caption{Diagram of the proposed algorithm. In this work, we use QMDP for the POMDP solver. Belief updates and QMDP solving are performed at a much faster rate than the computation of local NE.}
    \label{fig:alg:diagram}
    \vspace{-0.9\baselineskip}
\end{figure*} 
\subsection{POMDPs: Better Approximations for MaxEnt NE}
The main problem with LQ approximations for MaxEnt NE is that they work poorly when the non-ego agents' policies are not well approximated by a unimodal Gaussian. In the previous example, both LQ approximations alone are terrible approximations of true MaxEnt policy for \textcolor{red1}{$\pi^\two$} (\cref{fig:example}).
On the other hand, if we use a more complex distribution such as a mixture model to model the optimal policy for each timestep, the number of mixture components will grow exponentially in the time horizon and is not computationally tractable.

To resolve this issue, we consider the \add{following reformulation of the problem} proposed in \cite{so2022multimodal}.
Taking inspirations from bounded rationality, we assume the non-ego agents are only capable of solving MaxEnt NE with LQ structure, but the ego agent does not know which LQ approximation is used.
To model this, consider a set of $M$ different LQ NE $\{ (\vX^{(z)}, \vU^{(z)})_{z=1}^Z \}$, and let $z \in \{ 1, \dots, Z \}$ be a discrete latent random variable that determines which approximation is used by the non-ego agents. We now consider an extension of the MaxEnt Dynamic Game \eqref{eq:maxent_nash_def} where all non-ego agents have full knowledge of the value of $z$ and are playing the corresponding optimal policy $\pi^{(z)}$. However, the non-ego agents incorrectly assume the ego agent knows what the true mode is. This is now a \textit{dynamic game with incomplete information} --- the ego agent only has a \textit{belief} $b$ of what the true game is but does not know what the true dynamics are \cite{maschler2013game, so2022multimodal}. As shown in \cite{so2022multimodal}, the problem of computing the \textit{Bayesian equilibrium} \cite[Chapter 10]{maschler2013game} is equivalent to solving the following POMDP problem with state $\tilde{\vx} \coloneqq [\vx, b]$ (where we assume the ego agent is agent $\one$)
\begin{equation} \label{eq:pomdp_problem}
V_t^\one(\tilde{\vx}_t) = \inf_{\pi^\one}
\Eb_{z \sim b_t} \Eb_{\vu_t}\left[
    l^\one(\vx_t, \vu_t) + V_{t+1}^\one(\tilde{\vx}_{t+1}) \right],
\end{equation}
where $\vu^{\notone} \sim \vpi^{\notone, z}$ and
the dynamics of the belief $b$ follow the Bayesian filtering update law
\begin{equation} \label{eq:bayesian_filtering}
    b_{t+1}(z) \propto \pi^{\notone, z}_t(\vu^{\notone}_t | z) \, b_{t}(z).
\end{equation}
Since computing $b_t$ requires $\pi^{\notone,z}_{t-1}$, the LQ NE is solved from $\vx_{t-1}$.
We follow \cite{so2022multimodal} and choose the prior belief $b_0$ as the MaxEnt distribution over the sum of agent value functions
\begin{equation}
    b_0 = \inf_b \Eb_{z \sim b}\left[ \sum_{i=1}^N V^{i, z}(\vx_0) \right] - \frac{1}{\beta} H[b].
\end{equation}

\begin{remark}
Unlike the usual POMDP setup in literature with discrete state and action spaces \cite{kochenderfer2015decision,hauskrecht2000value,littman1995learning}, the above setup can be considered a POMDP with \textit{hybrid} states $[\vx, z]$ ($\vx$ is continuous but $z$ is discrete), infinite-dimensional controls $\pi^\one$ and continuous observations $\vx_t$. Since $\vx$ is fully observable, the belief $\tilde{b}$ is \textit{degenerate} on the $\vx$ coordinates.
\end{remark}

\subsection{MPOGames}
Leveraging this POMDP reformulation, we now present the MPOGames algorithm.
First, we solve for a set of different LQ MaxEnt NE, which can be done in parallel.
Since we assume each non-ego agent has picked one of these local NE and executes $\vpi^{\notone, z}$, we perform Bayesian inference using \eqref{eq:bayesian_filtering} to form a belief $b$ over $z$. This belief is used to solve a POMDP to give an optimal policy $\pi^{\one}$ for the ego agent. In practice, only the mean ego policy is used. A diagram of MPOGames is summarized in \cref{fig:alg:diagram}.
Intuitively, this reformulation improves the fidelity of $\pi^{\notone}$ from a single to a mixture of Gaussians. This enables a more accurate expectation over non-ego agents' policies $\pi^{\notone}$ in \eqref{eq:maxent_q}, which is key to the behavior of the MaxEnt solution.

Returning to the toy example, MPOGames recovers similar hedging behavior to the exact MaxEnt case (ego P$\one$).

\noindent\textbf{MPOGames: }
The value functions for P$\one$
take on the values $V^{\one, (1)} = 0.17$ and $V^{\one, (2)} = 0.09$.
Consequently, for $\epsilon = 0.1$,
\begin{equation}
    b_0 = [0.52, 0.48], \quad \textcolor{blue1}{\pi^{\one}} = \mathcal{N}\left(0.0, 0.5\right),
\end{equation}
which displays similar hedging behavior as the exact MaxEnt NE solution (see \cref{fig:example} left).

\subsection{Approximate Solutions for POMDPs}
Note that MPOGames does not prescribe a POMDP solver.
However, exact solutions for POMDPs are generally intractable \cite{kochenderfer2015decision}.
Two popular approximations for solving POMDPs are $Q_{\text{MDP}}$ \cite{littman1995learning} and fast informed bound (FIB) \cite{hauskrecht2000value}.
$Q_{\text{MDP}}$ has been used successfully in practice \cite{hauser2011online} and 
performs well when the particular action has little impact on the reduction in state uncertainty \cite{kochenderfer2015decision}.
It can be shown that when the state $\vx_t$ of all agents is directly observable, the $Q_{\text{MDP}}$ and FIB approximations are identical and equal to the MaxEnt NE value function. Moreover, if the value function $V$ for each mode is known, then the $Q_{\text{MDP}}$ can be computed easily.

An alternative approach for solving POMDPs is to use point-based methods which yield more accurate solutions \cite{kochenderfer2015decision}. Instead of changing the information structure (i.e., assuming the state is known), these methods work with the full value function on beliefs. 
For example, \cite{qiu2020latent,bernardini2012StabilizingModel,mesbah2018StochasticModel} consider a scenario tree based structure for solving the POMDP.
However, these methods still scale exponentially in the time horizon and have not been demonstrated to run in real-time in existing works \cite{qiu2020latent,hu2022active}. In comparison, QMDP with known $V$ has constant complexity in the time horizon.

In this work, we implement MPOGames using the QMDP approximation for its effectiveness at low computational costs. While the value function $V^{\one, z}$ for each mode is not known exactly, we take the quadratic approximation of $V$ obtained from solving the (feedback) NE for each mode.
The POMDP is solved with QMDP using the new belief and value functions, i.e.,
\begin{equation}
    \min_{\pi^{\one}_t} \Eb_{u^{\one}_t} \Eb_{z \sim b_t} \left[ l^{\one, z} + V^{\one, z}_{t+1}(\vx^{z}_{t+1}) \right],
\end{equation}
and the resulting solution is sent to a lower level trajectory tracker. Due to the efficiency of QMDP, we are able to solve it much faster than we can solve for local NE. Hence, we decouple the two parts in our algorithm to enable more responsive belief updates. The low level trajectory tracking is also decoupled to enable fast disturbance rejection from model errors in practice. The three decoupled nodes and their respective runtimes are presented in \cref{fig:alg:diagram}. 

\section{Simulation experiments}
\subsection{Autonomous Vehicle Merge} \label{sec:sim:av_merge}
\begin{figure*}
    \centering
    \includegraphics[width=0.999\linewidth]{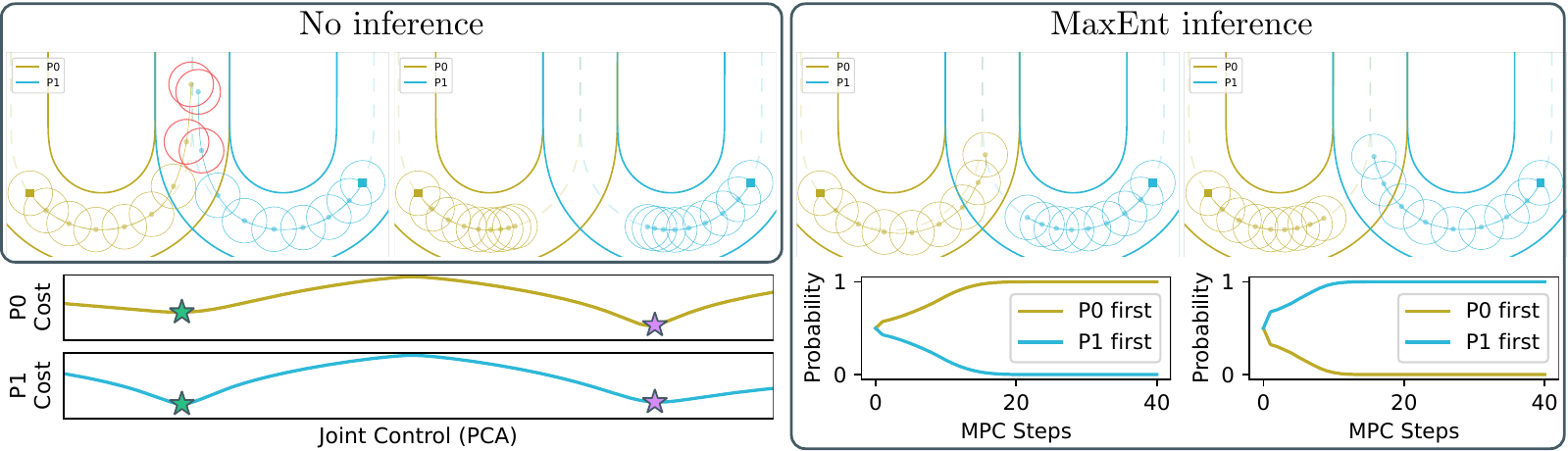}
    \caption{
        \textbf{Top Left, Top Right}: Simulation snapshots of a merge scenario. The ego agent is on the left and in \textbf{\textcolor{figgold}{gold}}. The circles denote the radius for each robot. \textbf{\textcolor{figred}{Red}} circles indicate collision.
        Being unable to consider multiple modes can lead to catastrophic failures such as collisions or freezing if the inferred mode is incorrect.
        \textbf{Bottom Left}: Visualization of the two local Nash equilibria for this problem.
        \textbf{Bottom Right}: Inferred probabilities for each mode via MaxEnt inference.
    }
    \label{fig:sim:noinf}
    \vspace{-0.6\baselineskip}
\end{figure*} We consider a merge scenario with two agents merging into the same lane while respecting track boundaries and avoiding collisions and denote the ego agent by P0.

Agents are modeled as kinematic bicycles. To easily handle both types of state constraints, we consider a \textit{hybrid} Cartesian-Curvilinear state representation \add{similar to \cite{zhu2022sequential}}. Let $x_i = [p_x, p_y, v, \theta, \zeta, s, n, \xi] \in \Rb^8$ and $\ui = [\dd{\zeta}, a]$ denote the state and control for agent $i$. The tracks are represented as a constant width lane around a centerline. Both the signed curvature $\kappa$ and the Cartesian coordinates of the centerline are represented as a cubic spline in the arclength $s$.

The cost for each agent consists of a quadratic cost for staying close to the centerline, maintaining a target velocity, and a soft constraint for staying inside the road boundaries. We handle collisions by treating them as soft-costs using a quadratic cost function on the Cartesian coordinates. Since this is concave in the position, there are two local NE --- either the ego agent goes first, or the non-ego agent goes first (\cref{fig:sim:noinf}). Consequently, we do not know which NE the non-ego agent is considering and must perform inference.

We consider the following game-theoretic methods in this scenario:
\begin{enumerate}
    \item (\ttNoInf) No inference is done for the non-ego agent's considered mode, and we assume that the non-ego agent considers the same mode as the one we are considering. This is the approach taken by methods such \cite{fridovich2020efficient,cleac2019algames} which only consider a single local NE. 
    \item (\ttML) Inference is done. However, instead of solving the POMDP, the controls corresponding to the maximum-likelihood (ML) mode are used as in \cite{peters2020inference}.
    \item (\ttQMDP) Inference is done, and the resulting POMDP is solved using the QMDP approximation.
\end{enumerate}
Although both \texttt{ML} and \texttt{MPOGames} perform inference, \texttt{ML} naively takes the most probable mode and ignores the rest. If the inferred probabilities are $b = [0.5 + \epsilon, 0.5 - \epsilon]$ for small $\epsilon$, \texttt{ML} applies the policy for the first mode, while \texttt{MPOGames} will consider both modes almost equally.

\begin{figure}
    \centering
    \includegraphics[width=0.999\linewidth]{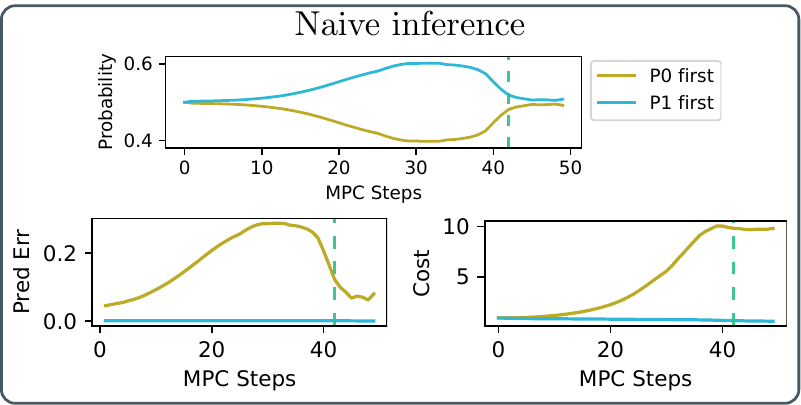}
    \caption{
        In contrast to the MaxEnt probabilities in \cref{fig:sim:noinf}, naive inference results in both modes having \textit{similar probabilities} near the end (eg. dashed line) due to both modes having similar prefixes at the merge point, despite the wrong mode having much higher cost.
\textbf{Top}: Inferred probabilities for each mode via MaxEnt inference.
        \textbf{Left}: Prediction error for each mode.
        \textbf{Right}: Cost of each predicted mode.
}
    \label{fig:sim:naive_inf}
    \vspace{-\baselineskip}
\end{figure}

Each method is run in a receding-horizon fashion with the results shown in \cref{fig:sim:noinf} and \cref{fig:sim:naive_inf}.
While \texttt{NoInf} performs well when the considered mode is the same as the non-ego agent, it fails when the mode is different --- either both agents believe the other agent will yield and crash into each other, or both agents yield resulting in both agents stopping. 
Despite modeling the response of other agents, the ``freezing'' behavior here happens because
the predicted response is incorrect.
Methods that perform inference (\texttt{ML}, \texttt{MPOGames}) infer the correct mode in the noiseless case and successfully perform the merge. Note that the likelihood of the wrong hypothesis when using \texttt{ML} actually \textit{increases} at the end of the trajectory in \cref{fig:sim:naive_inf}. This occurs because the trajectory prefix of both modes becomes increasingly similar at the merge point despite the cost of the wrong mode being much larger.

\begin{figure}
    \centering
\includegraphics[width=0.999\columnwidth]{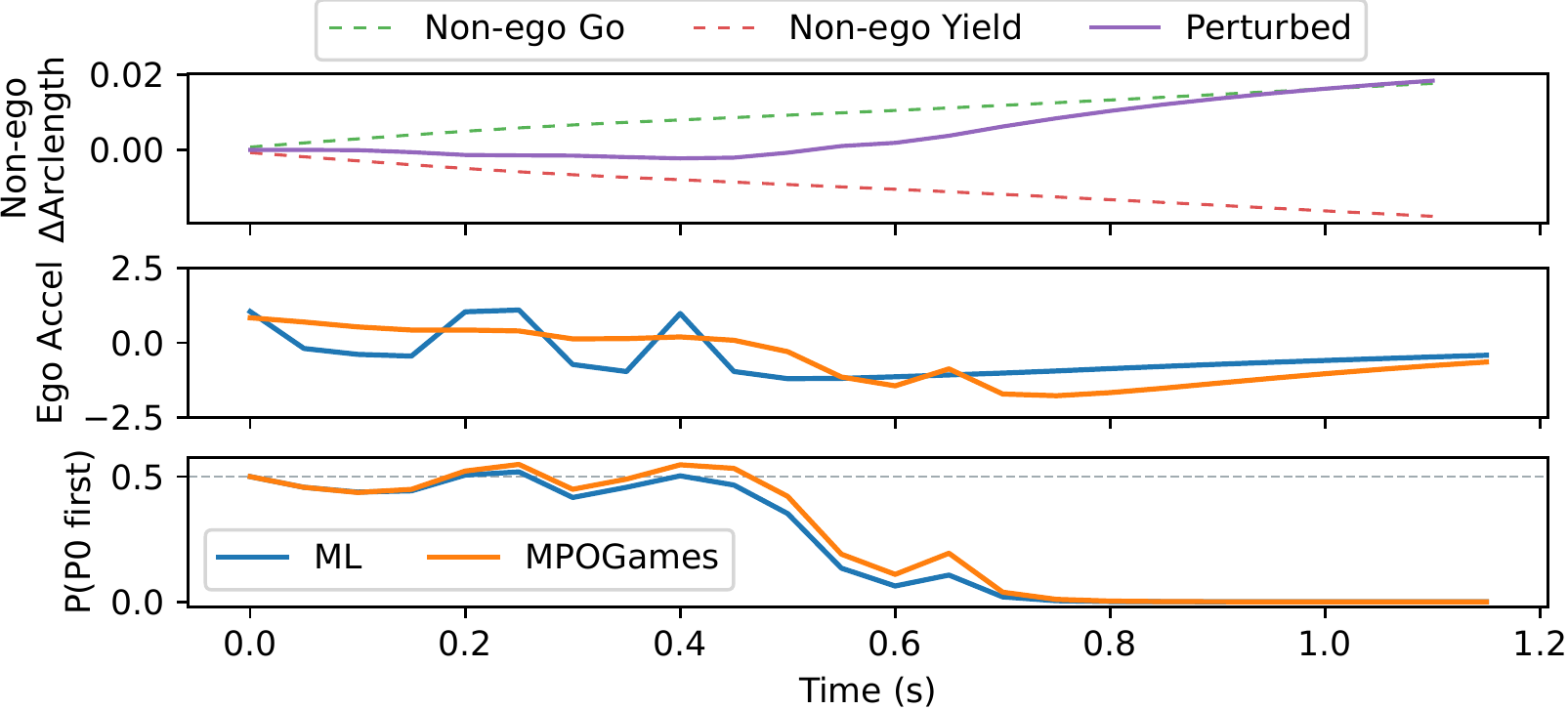}
    \caption{Comparison between the \texttt{ML} and \texttt{MPOGames} methods when the non-ego agent uses suboptimal controls.
    The discontinuous nature of ML solution means that small perturbations in the non-ego agent's controls (\textbf{top}) results in a discontinuous, jerky ego control (\textbf{center}) when the uncertainty oscillates around $0.5$ (\textbf{bottom}).
    On the other hand, \texttt{MPOGames} is a continuous function of the belief with smoother controls.
}
    \label{fig:sim:compare_ml_qmdp}
    \vspace{-1.4\baselineskip}
\end{figure}

We now consider the case where the non-ego agent's plans are not perfectly rational and perturb the non-ego agent's controls slightly such that there is large uncertainty regarding the non-ego agent's mode (\cref{fig:sim:compare_ml_qmdp}).
The high uncertainty of the true mode means that the maximum-likelihood mode oscillates between the two modes. Consequently, the discontinuity of \texttt{ML} results in a discontinuous, jerky ego control. In comparison, \texttt{MPOGames} hedges between the two possibilities and outputs a smooth, continuous control.

\noindent\textbf{Choice of inference technique. }
Additionally, we investigate the difference between a
naive method of inference via a user-specified Gaussian measurement model \cite{peters2020inference} and the MaxEnt inference presented in this paper. The results for naive inference are shown in \cref{fig:sim:naive_inf} (cf. \cref{fig:sim:noinf}).
Despite inferring the correct mode, this probability actually \textit{decreases} towards the end despite larger costs.

\noindent\textbf{Implementation Details.}
The symmetric structure of the coupling costs in the problems we consider makes this a potential game. Consequently, we solve for the (feedback) NE by reformulating the problem as an optimal control problem in a manner similar to \cite{kavuncu2021potential,bhatt2022efficient}.
However, we note that the NE can be solved with other ``backends'' such as iLQGames \cite{fridovich2020efficient}, ALGames \cite{cleac2019algames} or SQP \cite{zhu2022sequential}, with the value function and policy recovered by performing an additional backwards pass of iLQGames.
Next, although the belief update \eqref{eq:bayesian_filtering} requires access to the control inputs of other players, this information may not be directly available. Instead, we perform a least-squares estimate of $\vu_t^{\notone}$ using $\vx_t$ and $\vx_{t+1}$.
We also assume prior knowledge on how to find a comprehensive set of NE. In this work, this is done using a heuristic initialization process and fixed to two NE. Determining how many NE exist and how to find this set of NE is beyond the scope of this work.
We use CasADi \cite{andersson2019casadi} and acados \cite{verschueren2022acados} to solve the resulting optimal control problem with $\Delta t=\SI{0.05}{\second}$ and horizon $T=60$.

\section{Hardware experiments} \label{sec:hardware}

Finally, we validate
our approach on hardware in a merge scenario.
We use the TRIKart \cite{dimitrov2022rapid},
a modified version of the F1Tenth platform \cite{o2020f1tenth}. Optitrack \cite{optitrack} is used for localization.

\subsection{No Perturbations}
We first investigate how each method performs in the cases where the non-ego agent is cooperative (\texttt{Yield}), selfish (\texttt{NoYield}), or using the \texttt{ML} or \texttt{MPOGames} strategies.
The first two cases (\texttt{Yield}, \texttt{NoYield}) are realized by performing \texttt{NoInf} with an initialization strategy that is biased towards yielding or not yielding respectively. The success rate of merging over ten trials is shown in \cref{tab:hw:matrix}, where success is defined loosely as both agents reaching the center lane without colliding or stopping.

\begin{figure}
    \centering
\includegraphics[trim=0 250 220 240,clip,width=0.88\columnwidth]{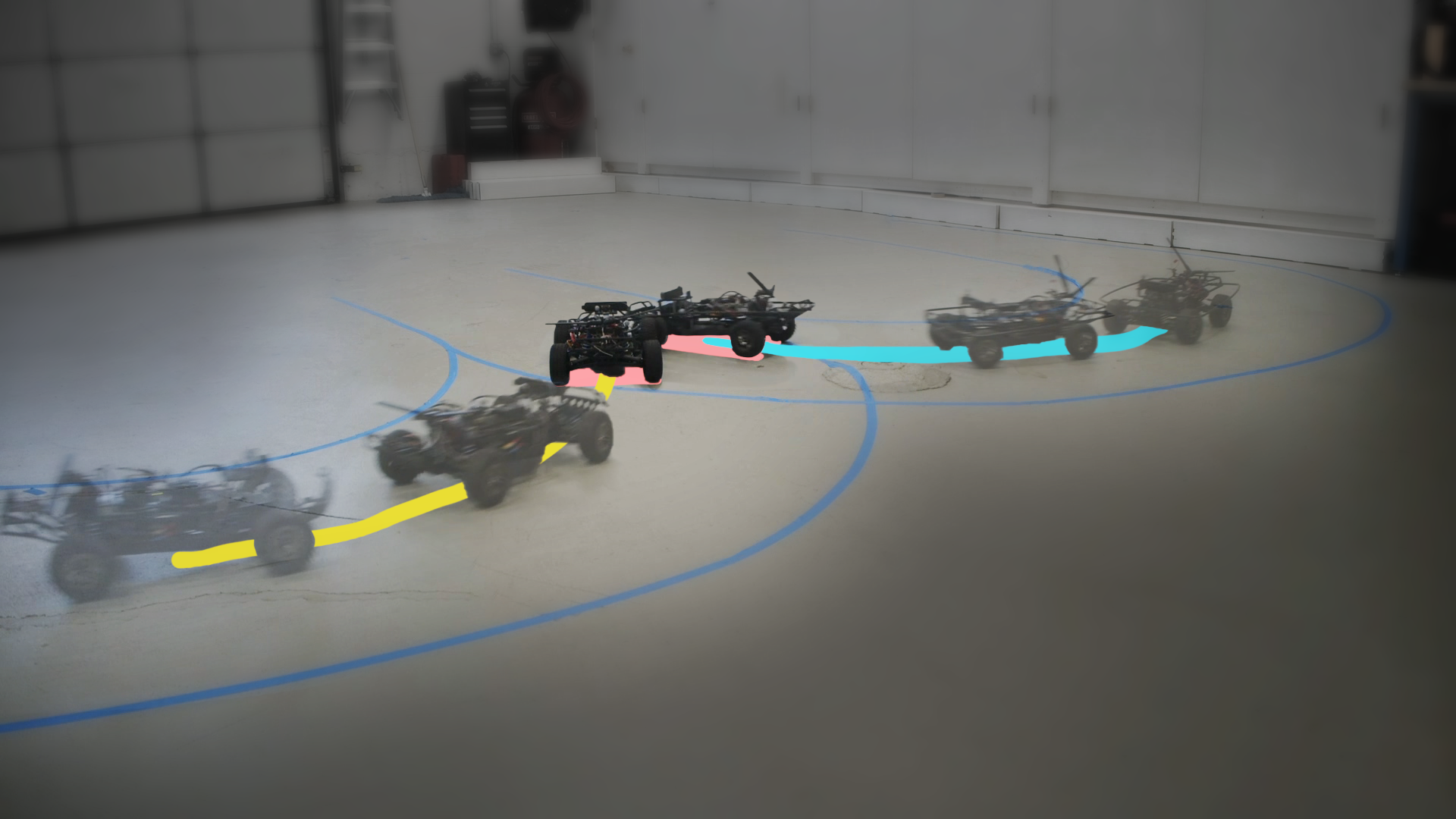}
    \caption{\texttt{NoYield-NoYield} setup.
Collisions occur when agents do not agree on the same local NE. Left is ego agent.}
    \label{fig:hw:collide}
    \vspace*{-1.0\baselineskip}
\end{figure}

\noindent\textbf{\texttt{Yield}, \texttt{NoYield} --- Mismatch of Local NE: }
As predicted from \cref{sec:sim:av_merge}, failures occur when there is a mismatch in local NE between the two ages (i.e., \texttt{Yield-Yield} and \texttt{NoYield-NoYield}). In the \texttt{Yield-Yield} case, all failures arose from both agents coming to a complete stop before the merge point. Due to noise, one of the agents eventually merges, but this only happens after a long delay.
The two successes in the \texttt{NoYield-NoYield} example arose when one car was much closer to the merge point than the other car due to hardware differences, resolving the ambiguity. However, \texttt{NoYield-NoYield} was also the only combination that resulted in collisions (see \cref{fig:hw:collide}).

\begin{figure}[t]
    \centering
    \includegraphics[width=0.95\columnwidth]{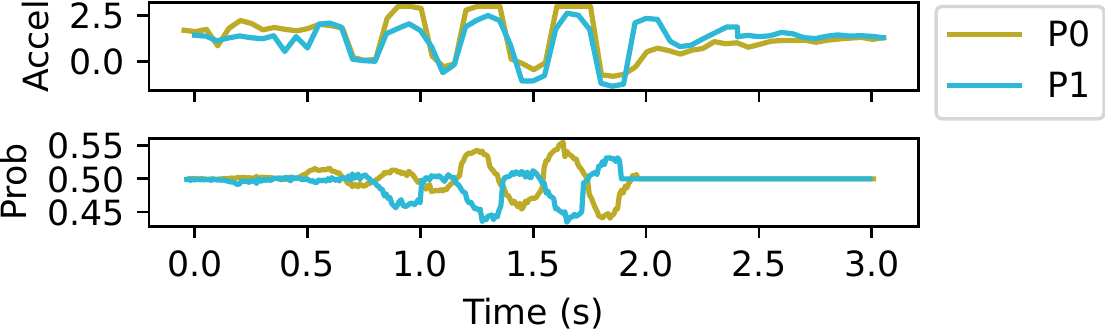}
    \vspace*{-0.2\baselineskip}
    \caption{Results on the scale car hardware when both agents use \texttt{ML}.
The processing latency leads to an unstable, oscillatory system.
\textbf{Top:} Output accelerations. \textbf{Bottom:} inferred probabilities of $P0$ (Ego) going first. }
    \label{fig:hw:mlml_osc}
\vspace{1.2\baselineskip}
\includegraphics[width=0.95\columnwidth]{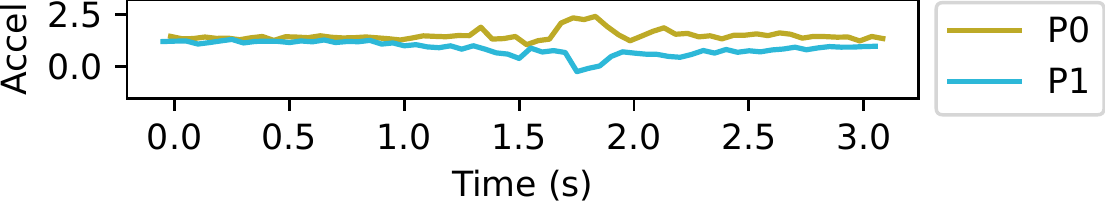}
    \vspace*{-0.2\baselineskip}
    \caption{Same setup as \cref{fig:hw:mlml_osc}, but both agents use \texttt{MPOGames}.}
    \label{fig:hw:qmdpqmdp_osc}
\vspace{1.2\baselineskip}
\includegraphics[width=0.95\columnwidth]{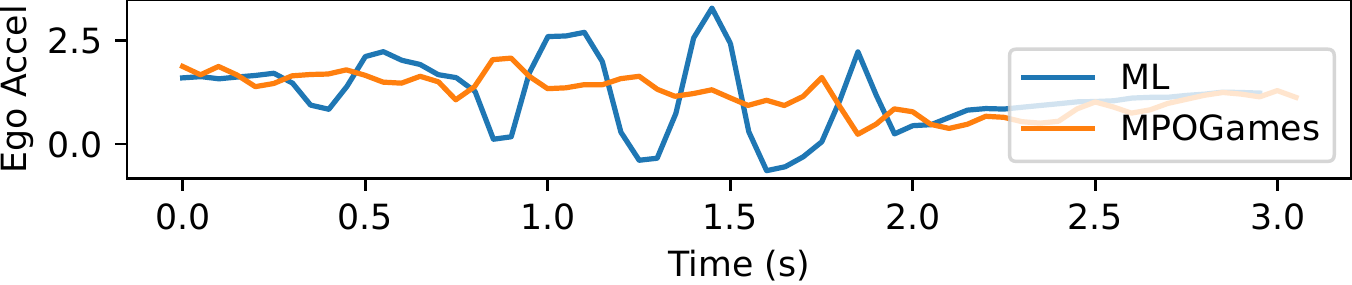}
    \vspace*{-0.2\baselineskip}
    \caption{Comparison of accelerations between \texttt{ML} and \texttt{MPOGames} \textit{on hardware} when the non-ego agent applies high uncertainty controls. Note the oscillation for \texttt{ML}.}
    \label{fig:hw:cmp_osc}
    \vspace*{-1.0\baselineskip}
\end{figure}

\noindent\textbf{\texttt{ML} --- Vulnerablility to system latency: }
The simulation experiments in \cref{sec:sim:av_merge} make the standard assumption that there is no delay between receiving observations and outputting the controls. However, this is not true in the real world. In the \texttt{ML-ML} setup, although all trials result in successful merges, both agents exhibit large oscillatory controls (\cref{fig:hw:mlml_osc}). This behavior is inherent to the discontinuous behavior of \texttt{ML}.
Despite having the same system latency, \texttt{MPOGames-MPOGames} handles this situation more gracefully due to accounting for uncertainty (\cref{fig:hw:qmdpqmdp_osc}).

\begin{table}
\centering
\caption{
    Merge success rate on the scale car platform over ten trials from the same initial conditions.
}
\vspace*{-0.5\baselineskip}
\label{tab:hw:matrix}
\resizebox{0.9\columnwidth}{!}{\begin{tabular}{cr
    S[table-format=1.1]
    S[table-format=1.1]
    S[table-format=1.1]
    S[table-format=1.1]
    }
    \toprule
    & & \multicolumn{4}{c}{Non-ego Agent} \\ \cmidrule{3-6}
    & & \ttYield & \ttNoYield & \ttML & \ttQMDP \\ \midrule
    \multirow{4}{*}{Ego Agent}
    & \ttYield  & \cellcolor{palered}{0.0} \\
    & \ttNoYield& \cellcolor{palegreen}{1.0} & \cellcolor{palered}{0.2} \\
    & \ttML     & \cellcolor{palegreen}{1.0} & \cellcolor{palegreen}{1.0} & \cellcolor{palegreen}{1.0} \\
    & \ttQMDP   & \cellcolor{palegreen}{1.0} & \cellcolor{palegreen}{1.0} & \cellcolor{palegreen}{1.0} & \cellcolor{palegreen}{1.0} \\
    \bottomrule
\end{tabular}
}
\vspace*{-0.2\baselineskip}
\end{table} 
  

\subsection{Controls with High Uncertainty}
\vspace*{-0.1\baselineskip}
Finally, we compare how each method performs when the non-ego agent has explicitly suboptimal controls. Specifically, we add sinusoidal noise to the non-ego agent's controls such that it is difficult to determine the non-ego agent's intentions. Here, we only compare \texttt{ML} and \texttt{MPOGames} and show the results in \cref{fig:hw:cmp_osc}. This is similar to the scenario in \cref{fig:sim:compare_ml_qmdp}, and indeed we see the same oscillations that \texttt{ML} displays on hardware. However, in this case, the non-ego agent is intentionally suboptimal.
 Again, \texttt{MPOGames} handles this uncertainty gracefully and correctly hedges against both NE.
 \section{Discussion}
\vspace*{-0.1\baselineskip}
\noindent \textbf{Summary: } We have proposed a MaxEnt game-theoretic framework for planning and inference in multi-agent situations with multiple local Nash equilibrium. Through simulation and hardware case studies, we have shown that MPOGames provides a real-time method for hedging against the uncertainty of other agents and improves performance compared to previous game-theoretic approaches.

\noindent \textbf{Limitations and future work:}
Using a GT planner for joint planning and prediction assumes the knowledge of the cost function and dynamics for all agents. However, this is rarely true and consequently must be \textit{estimated} using techniques such as inverse reinforcement learning \cite{ziebart2008maximum,cleac2020lucidgames}.
While the proposed reformulation into a POMDP relies on the key assumption that the non-ego agents know the true mode, we demonstrate that this restriction may be relaxed (e.g., \texttt{MPOGames-MPOGames} in \cref{sec:hardware}).
Also, our framework assumes that non-ego agents are only capable of solving LQ games. Relaxing this assumption to model more intelligent non-ego agents is left as future work.
Moreover, we have assumed prior knowledge on finding all NE. Doing so efficiently for general games is left as future work.
Finally, the current work assumes that there are a discrete set of modes, but there exist scenarios when there are a \textit{continuum} of NE. We leave this direction as future work. 
\FloatBarrier

\bibliographystyle{IEEEtran}
\bibliography{IEEEabrv,references.bib}

\end{document}